**Solid-liquid equilibria and excess enthalpies in binary mixtures of acetophenone with some aliphatic amides**

**Ana Cobos Huerga [a], Patryk Sikorski [b], Juan Antonio Gonzalez [a], Tadeusz Hofman[b,*]**

[a] Universidad de Valladolid, Dpto Física Aplicada, Facultad de Ciencias, Pase de Belén, 47011 Valladolid, Spain
[b] Warsaw University of Technology, Faculty of Chemistry, ul. Noakowskiego 3, 00-664 Warszawa, Poland

**ABSTRACT**

Solid-liquid equilibria for the binary systems of acetophenone and {N-methylformamide, or N,N-dimethylformamide, or N,N-dimethylacetamide, or N-methyl-2-pyrrolidone} were determined by the cloud-point technique. For the same systems, excess enthalpies were measured at 293.15 K and 318.15 K by the titration calorimetry. Both types of data were correlated by the Redlich-Kister equation. The description of the solid-liquid equilibria incorporated measured excess enthalpies. The results of the prediction performed by the modified UNIFAC model were compared with the experimental data. The observed trends and differences between systems were discussed.

**Keywords**: Amide, Acetophenone, Solid-liquid equilibrium, Excess enthalpy, UNIFAC

**1. Introduction**

Thermodynamic properties of mixtures with strong interactions still are not sufficiently understood. Particularly important are those which exhibit a chemical-like character what manifests in many peculiar properties sometimes called as anomalies. The hydrogen-bond interactions and their and characteristic temperature dependence is a good example. On the other hand, "ordinary" interactions are only slightly dependent on temperature. It is not clear if strong dipole-dipole interactions may be treated as the chemical-like ones. Special difficulties in interpretations can bear functional groups possessing donor and acceptor properties which may be responsible for directional interactions.

We have undertaken systematic studies concerning systems composed of compounds containing amide and carbonyl groups in which strong and diverse molecular interactions appear. Acetophenone is a compound frequently used in the industry, mainly as a precursor in some polymerization processes. It is also a substrate in pharmaceutical synthesis and because of its non-toxicity and strong flavor is used as a component in fragrances. Its carbonyl group is mainly responsible for strong molecular interactions because of its high dipole moment. Additionally, the aromatic ring also may bring a significant impact. The second component is a simple aliphatic amide (N-methylformamide - MF, N,N-dimethylformamide - DMF, N,N-dimethylacetamide - DMA) or lactam (N-methyl-2-pyrrolidone – NMP). All are frequently used as polar solvents possessing very high values of dielectric constant. It may be expected that the extremely high dielectric constant of N-methylformamide, equal to 189.0 at 293.2 K [1], strongly influences properties of its mixtures.

*Corresponding author at: Warsaw University of Technology, Faculty of Chemistry, ul. Noakowskiego 3, 00-664 Warszawa, Poland.
*E-mail address*: hof@ch.pw.edu.pl

These systems are also interested from the perspective of the solution-of-groups approach [2]. The crucial step in any group contribution is a definition of a functional group. The number of unique groups should be as low as possible. On the other hand they should be universal, i.e. their properties should be transferable and independent of the environment, that is adjacent groups. The possible definition a group vary between single atoms up to the group being equivalent to whole molecule [3]. The amide group usually is treated as a whole. However, it can be also interpreted as a superposition of amine and carbonyl group. In the mixture with a carbonyl compound, mutual interactions between carbonyl groups existing in both components can suppress deviations from ideality.

The thermodynamic data reported in the literature concerning the systems with acetophenone and the above mentioned amides are rare and mainly include volumetric and viscosity data. Vittal Prasad et al. measured isobaric vapor-liquid equilibria for the mixture of acetophenone with DMF at 96 kPa [4]. For the mixture with DMA, the excess enthalpy data at 298.15 K by Sekhar et al. are available [5]. Sedlakova et al. measured liquid-liquid and solid-liquid equilibria for the system formed with acetophenone and formamide[6]. The volumetric and viscosimetric data are more numerous and cover excess volumes or mixture densities, and viscosities for the following systems: acetophenone: + formamide at 298-318 K [7], +NMF at 303-318 K [8], + DMF at 293-313 K [9], + M-methylacetamide at 308.15 K [10] ,and + DMF (only excess volumes) at 308.15 K [11].

The main aim of the experimental part of this paper is to provide additional data which would enable characterizing and understanding of the molecular interactions occurring in these systems. The solid-liquid equilibrium and excess enthalpies data were measured for the binary mixtures of acetophenone with NMF, or DMF, or DMA and or NMP. From the latter data the excess heat capacities were estimated.

## 2. Materials

Acetophenone, N-methylformamide, N,N-dimethylformamide, N,N-dimethylacetamide and N-methyl-2-pyrrolidone were purchased in Sigma-Aldrich company. Purity of N-methylformamide confirmed by the GC was 99.6 %. Detailed information is given in Table 1.

Table 1. The compounds used in measurements

| Compound | CAS no. | Purity[a]/ % | Remarks |
|---|---|---|---|
| Acetophenone | 98-86-2 | 99 | |
| N-methylformamide | 123-39-7 | 99 | |
| N,N-dimethylformamide | 68-12-2 | 99.8 | anhydrous |
| N,N-dimethylacetamide | 127-19-5 | 99.8 | anhydrous |
| N-methyl-2-pyrrolidone | 872-50-4 | 99.5 | anhydrous |

[a]Declared by producer

## 3. Experimental

### 3.1. Solid-liquid equilibria

Majority of the solid-liquid equilibria data were measured using cloud-point technique and the apparatus was described in details elsewhere [12]. The very old idea of the experimental technique usually is being traced back to the XIXth century [13]. The schematic drawing of the apparatus is given in Fig. 1.

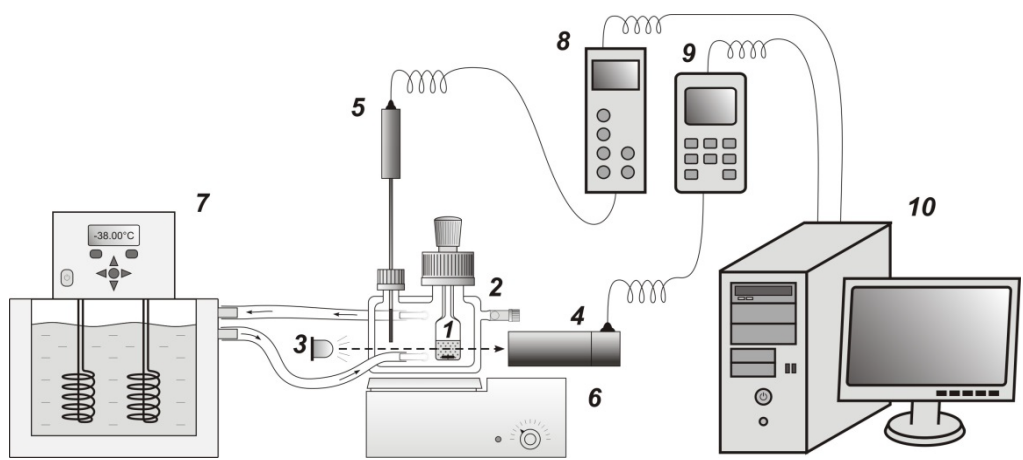

Fig. 1
The cloud-point technique apparatus: (1) equilibrium cell, (2) insulated vessel; (3) source of light; (4) probe measuring the intensity of light; (5) temperature probe; (6) magnetic stirrer; (7) cryostat; (9) detector of light; (10) CPU.

The experiment begins at sufficiently low temperature at which, with known masses of both components, a solid phase appears. Then, with a steady vigorous mixing, the temperature is slowly increased while the intensity of light beam crossing the mixture is monitored. As during this process an amount of crystals decreases, the luminance is increasing up to the a fixed level which indicates full solubility. The observed function of the light intensity with respect to temperature enables to determine solubility temperatures which correspond to the break in this dependence. For the measurements close to the solubility temperature, the heating rate did not exceed xxxx. The composition of the saturated solution was corrected due to evaporation of solvent into accessible part of the vessel. The usual amounts of the mixtures were between 0.5-2 cm$^3$ while the total volume of the measuring cell was about 4 cm$^3$. Standard uncertainty of temperature was estimated to be about 0.5 K.

The minimum solubility temperature which could be detected this way was determined by the cryostat efficiency and was about -30 $^o$C. At such low temperatures we encountered serious problems to crystalize samples because of extensive subcooling. It was necessary to decrease the solution temperature of about 30-40 K below the solubility temperature what was achieved by immersing a sample in the mixture of dry ice with acetone.

The solubility temperatures below the detection limits were determined by the Differential Scanning Calorimetry using the Mettler Toledo DSC1 STAR$^®$ apparatus. The mixtures with a well-defined composition were prepared by weighing with the RADWAG analytical balance. The samples of about 0.5-1.0 g were measured with repeatability of ±0.04 mg. An amount of 8-15 mg of the mixture was injected into an aluminum box with a pin and an empty vessel was used as a reference sample. The samples were kept for 10 minutes at 100°C which was achieved by means of liquid nitrogen and next they were heated with the rate of 1 K·min$^{-1}$. The STAR software was used to analyze data and determine eutectic and solubility temperatures. The former was derived from onset temperature while the latter one from the peak minimum. The standard expanded uncertainty of temperature was declared by the producer to be 0.1 K, although data scattering suggests higher value, at least 1 K. [add something on calibration]

The melting temperatures of pure compounds together with the corresponding enthalpies are given in Table 2. Measured solid-liquid equilibria are shown in Table 3. The data acquired by the DSC technique are written in italics. Two data sets systems (acetophenone + NMF and acetophenone + DMA) are also illustrated in Fig. 2 and 3.

Table 2. Melting temperatures and enthalpies of fusion of pure compounds

| | $T_m$/K | $\Delta H_m$/ kJ·mol$^{-1}$ | reference |
|---|---|---|---|
| Acetophenone | 292.7 | | [14] |
| | | 16.65 | [15] |
| | 293.0 | | [16] |
| | 291.7 | | this work |
| N-methylformamide | 270.6 | 10.44 | [17] |
| | 270.8 | | this work |
| N,N-dimethylformamide | 212.9 | 8.95 | [18] |
| | 212.1 | | this work |
| N,N-dimethylacetamide | 253.2 | 8.2 | [19] |
| | 254.2 | 10.2 | [18] |
| | 251.4 | 10.4 | [17] |
| | 252.7 | | this work |
| N-methyl-2-pyrrolidone | 248.5 | 18.1 | [19] |
| | 248.6 | | this work |

Table 3. Solid-liquid equilibria of the acetophenone (1) + {NMF, or DMF, or DMA, or NMP} systems. The data obtained by the DSC methods are written in italics.

| $x_1$ | $T$/K | phase | $\gamma_1$ | | $\gamma_2$ | |
|---|---|---|---|---|---|---|
| | | | exp | calc | exp | calc |
| Acetophenone (1) + N-methylformamide (2) | | | | | | |
| 0 | 270.8 | 2 | | | 1 | 1 |
| 0.0502 | 268.1 | 2 | | 3.098 | 1.005 | 1.002 |
| 0.0994 | 265.3 | 2 | | 2.812 | 1.008 | 1.010 |
| 0.1441 | 263.8 | 2 | | 2.567 | 1.033 | 1.024 |
| 0.1999 | 262.7 | 1 | 2.340 | 2.342 | | 1.043 |
| 0.2494 | 267.0 | 1 | 2.121 | 2.138 | | 1.069 |
| 0.3000 | 270.5 | 1 | 1.945 | 1.957 | | 1.104 |
| 0.3496 | 273.8 | 1 | 1.826 | 1.808 | | 1.144 |
| 0.3886 | 275.8 | 1 | 1.729 | 1.707 | | 1.181 |
| 0.4476 | 277.7 | 1 | 1.581 | 1.579 | | 1.247 |
| 0.5002 | 279.6 | 1 | 1.485 | 1.483 | | 1.315 |
| 0.5563 | 281.1 | 1 | 1.386 | 1.396 | | 1.403 |
| 0.6028 | 282.6 | 1 | 1.328 | 1.333 | | 1.493 |
| 0.6446 | 283.5 | 1 | 1.272 | 1.281 | | 1.591 |
| 0.6945 | 284.7 | 1 | 1.215 | 1.223 | | 1.743 |
| 0.7503 | 286.0 | 1 | 1.162 | 1.164 | | 1.980 |
| 0.8007 | 287.2 | 1 | 1.121 | 1.114 | | 2.294 |
| 0.8497 | 288.2 | 1 | 1.083 | 1.072 | | 2.754 |
| 0.9002 | 289.6 | 1 | 1.057 | 1.035 | | 3.506 |
| 0.9502 | 290.9 | 1 | 1.031 | 1.010 | | 4.763 |
| 1 | 291.7 | 1 | 1 | 1 | | |
| Acetophenone (1) + N,N-dimethylformamide (2) | | | | | | |
| 0 | 212.1 | 2 | | | | 1 |
| 0.0398 | 211.0 | 2 | | 0.543 | 1.014 | 0.996 |
| 0.0602 | 208.9 | 2 | | 0.587 | 0.985 | 0.992 |
| | 206.0 | 1+2 | | | | |
| 0.0803 | 206.2 | 1+2 | | | | |
| 0.1203 | 213.4 | 1 | 0.670 | 0.662 | | 0.983 |
| | 205.6 | 1+2 | | | | |
| 0.1518 | 206.4 | 1+2 | | | | |
| 0.2066 | 231.5 | 1 | 0.812 | 0.818 | | 0.949 |
| | 206.4 | 1+2 | | | | |
| 0.2511 | 205.3 | 1+2 | | | | |
| 0.2981 | 245.3 | 1 | 0.916 | 0.930 | | 0.915 |
| | 206.5 | 1+2 | | | | |
| 0.3004 | 245.5 | 1 | 0.915 | 0.932 | | 0.914 |
| | 207.5 | 1+2 | | | | |
| 0.30202 | 247.0 | 1 | 0.956 | 0.933 | | 0.913 |
| 0.3509 | 251.1 | 1 | 0.996 | 0.971 | | 0.898 |

| $x_1$ | $T$ | phase | | | | |
|---|---|---|---|---|---|---|
| | *206.2* | 1+2 | | | | |
| 0.35341 | 253.2 | 1 | 0.939 | 0.970 | | 0.899 |
| 0.39636 | 257.3 | 1 | 1.008 | 0.992 | | 0.888 |
| 0.44788 | 261.1 | 1 | 0.999 | 1.007 | | 0.881 |
| 0.49878 | 265.4 | 1 | 1.015 | 1.013 | | 0.878 |
| 0.55268 | 269.2 | 1 | 1.019 | 1.013 | | 0.880 |
| 0.60509 | 272.4 | 1 | 1.016 | 1.009 | | 0.886 |
| 0.65299 | 275.0 | 1 | 1.009 | 1.005 | | 0.893 |
| 0.70014 | 277.5 | 1 | 1.005 | 1.001 | | 0.902 |
| 0.75259 | 279.6 | 1 | 0.987 | 0.998 | | 0.912 |
| 0.79736 | 281.8 | 1 | 0.985 | 0.997 | | 0.918 |
| 0.84764 | 284.4 | 1 | 0.989 | 0.996 | | 0.920 |
| 0.90204 | 286.8 | 1 | 0.986 | 0.998 | | 0.913 |
| 0.94877 | 289.1 | 1 | 0.991 | 0.999 | | 0.897 |
| 1 | 291.7 | 1 | 1 | 1 | | |

### Acetophenone (1) + N,N-dimethylacetamide (2)

| $x_1$ | $T$ | phase | | | | |
|---|---|---|---|---|---|---|
| *0* | *252.7* | 2 | | | 1 | 1 |
| 0.04946 | 249.3 | 2 | | 0.699 | 0.985 | 0.998 |
| 0.09949 | 246.6 | 2 | | 0.750 | 0.985 | 0.991 |
| *0.1004* | *247.5* | 2 | | 0.751 | 1.004 | 0.991 |
| *0.1512* | *244.7* | 2 | | 0.802 | 1.005 | 0.980 |
| | *235.6* | 1+2 | | | | |
| *0.2019* | *239.6* | 2 | | 0.853 | 0.961 | 0.965 |
| 0.34946 | 247.6 | 1 | 0.843 | 0.846 | | 0.972 |
| 0.39872 | 253.3 | 1 | 0.886 | 0.879 | | 0.957 |
| *0.3998* | *235.2* | 1+2 | | | | |
| 0.44772 | 257.5 | 1 | 0.897 | 0.907 | | 0.942 |
| 0.50472 | 262.7 | 1 | 0.929 | 0.933 | | 0.924 |
| 0.54759 | 266.0 | 1 | 0.941 | 0.949 | | 0.912 |
| 0.59770 | 269.9 | 1 | 0.961 | 0.964 | | 0.898 |
| 0.64049 | 273.5 | 1 | 0.989 | 0.975 | | 0.887 |
| 0.70323 | 277.3 | 1 | 0.996 | 0.986 | | 0.873 |
| 0.74457 | 280.3 | 1 | 1.016 | 0.991 | | 0.865 |
| 0.80350 | 282.3 | 1 | 0.990 | 0.996 | | 0.856 |
| 0.85237 | 284.7 | 1 | 0.991 | 0.998 | | 0.851 |
| 0.90413 | 287.0 | 1 | 0.988 | 1.000 | | 0.848 |
| 0.95185 | 289.2 | 1 | 0.990 | 1.000 | | 0.848 |
| 1 | 291.7 | 1 | 1 | 1 | | |

### Acetophenone (1) + N-methyl-2-pyrrolidone (2)

| $x_1$ | $T$ | phase | | | | |
|---|---|---|---|---|---|---|
| *0* | *248.6* | 2 | | | 1 | 1 |
| *0.0507* | *246.8* | 2 | | 0.268 | 0.988 | 0.991 |
| *0.0964* | *245.2* | 2 | | 0.346 | 0.980 | 0.971 |
| | *229.7* | 1+2 | | | | |
| 0.09811 | 245.9 | 2 | | 0.349 | 1.007 | 0.970 |
| *0.1505* | *241.5* | 2 | | 0.451 | 0.910 | 0.933 |

| $x_1$ | $T$ | region | | | | |
|---|---|---|---|---|---|---|
|  | 229.7 | 1+2 |  |  |  |  |
| 0.2008 | 238.8 | 2 |  | 0.570 | 0.874 | 0.883 |
|  | 231.3 | 1+2 |  |  |  |  |
| 0.40405 | 248.2 | 1 | 0.743 | 0.737 | 0.809 |  |
| 0.44987 | 254.3 | 1 | 0.810 | 0.806 | 0.769 |  |
| 0.50509 | 260.5 | 1 | 0.870 | 0.871 | 0.728 |  |
| 0.54732 | 264.6 | 1 | 0.905 | 0.910 | 0.701 |  |
| 0.60399 | 269.5 | 1 | 0.941 | 0.948 | 0.672 |  |
| 0.65507 | 273.0 | 1 | 0.954 | 0.971 | 0.652 |  |
| 0.69511 | 276.4 | 1 | 0.984 | 0.984 | 0.641 |  |
| 0.75117 | 280.3 | 1 | 1.007 | 0.994 | 0.630 |  |
| 0.78624 | 281.9 | 1 | 1.002 | 0.998 | 0.626 |  |
| 0.85417 | 285.7 | 1 | 1.014 | 1.001 | 0.626 |  |
| 0.90226 | 288.4 | 1 | 1.025 | 1.001 | 0.631 |  |
| 0.95080 | 290.6 | 1 | 1.025 | 1.000 | 0.641 |  |
| 1 | 291.7 | 1 | 1 | 1 |  |  |

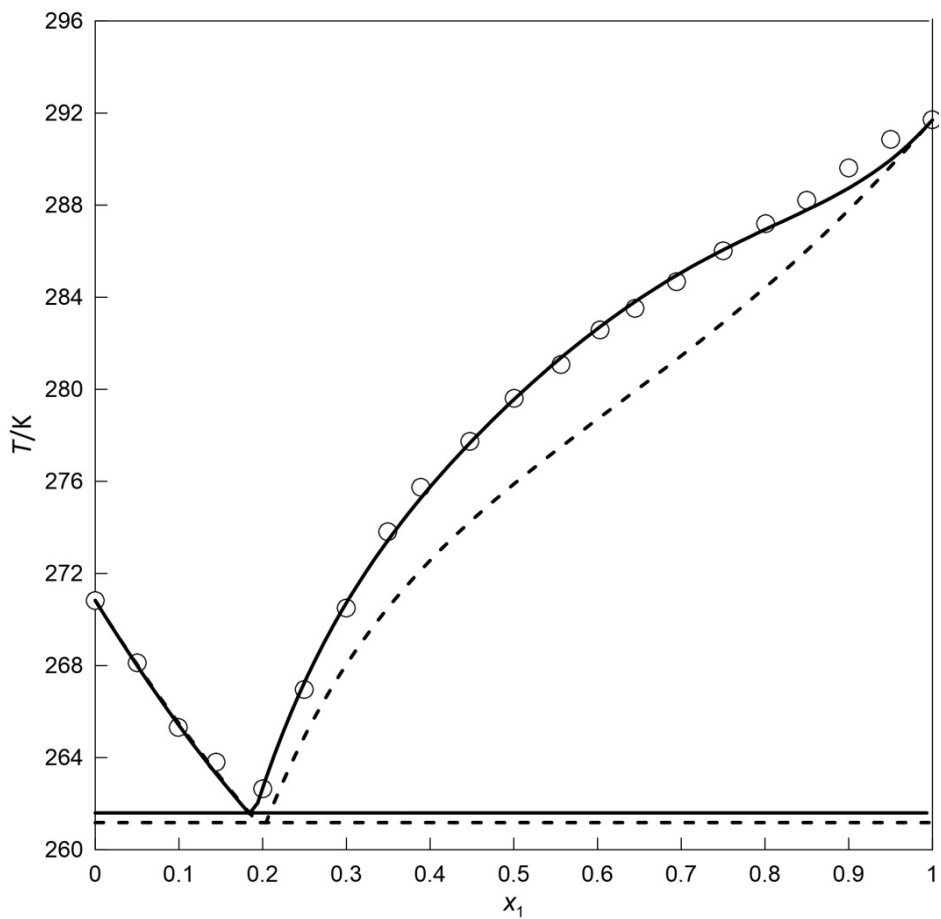

Fig. 2

Solid-liquid equilibria for the acetophenone + N-methylformamide system: (○) − experimental, solid line − calculated by eqn. (4) with the parameters given in Table 6, dashed line − predicted by the modified UNIFAC model.

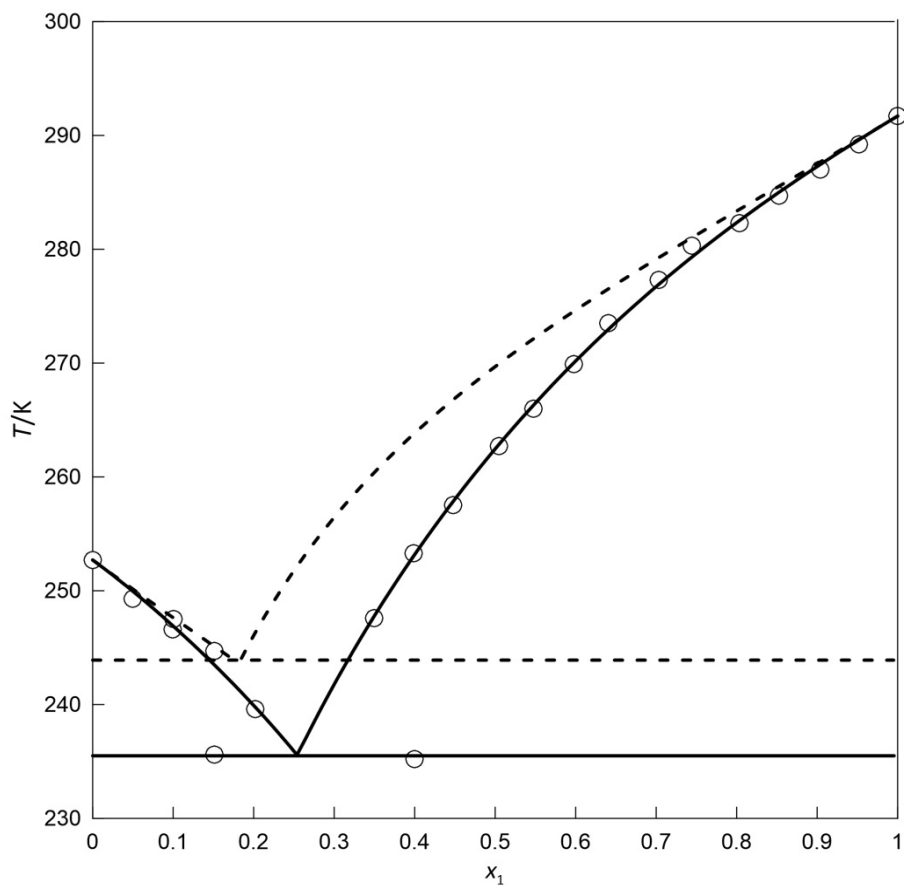

Fig. 3

Solid-liquid equilibria for the acetophenone + N,N-dimethylacetamide system: (○) – experimental, solid line – calculated by eqn. (4) with the parameters given in Table 6, dashed line – predicted by the modified UNIFAC model.

### 3.2. Excess Enthalpy

The excess enthalpies of mixing were determined by means of isothermal titration calorimetry. For this purpose the TAM III calorimeter was used. This apparatus uses the thermostatic oil bath to maintain the temperature with stability of ±100 μK. The measurements were conducted at 293.15 K and 308.15 K, with the temperature kept constant for 24 hours prior to the experiment.

In the beginning of a measurement about 0.5 mL of one component  was placed in a stainless steel ampule serving as the titration cell and the same amount was introduced into the reference cell. Both cells were placed in the termostatic oil bath. The equilibration of cells took a few hours. Next appropriate amounts of the second compound were injected using the syringe pump to provide the change in mole fractions of about 0.3. The mole fractions were determined volumetrically using previously measured densities of the compounds. The mixture of both compounds was stirred with a speed of 100 rpm. In the second series of experiments the second component was placed in the titration and reference cells while the first one was injected by the syringe.

The calorimeter measures differences between heat flows of the reference and sample cells, with uncertainty of about 0.2%.

Integration of the heat flow peaks enables us to calculate the molar excess enthalpy of mixing using the following equation which standard uncertainty is estimated as to be lower than 0.5%, for $i$th injection as follows:

$$H_i^E = \frac{\sum_{j=1}^i \delta q_j}{n_1 + \sum_{j=1}^i \Delta n_{2j}}$$

Where $\delta q_j$ is the heat effect during the jth injection, $n_1$ is the number of moles of compound 1 in the titration cell before the titration starts and $\Delta n_{2j}$ is the number of moles of compound 2 injected into titration cell during the $j$th tritration.

The measured enthalpies are given in Table 4 and two exemplary dependences (for the acetophenone + NMF and acetophenone + DMA) are shown graphically in Fig. 4 and 5. The excess enthalpies data for the former system are the solely measurements which can be compared with the previously reported literature data measured by Sekhar et al. at 298.15 K [5]. They deviate considerably from ours being approximately ten times more negative. In our opinion, an extent of the literature data does not agree with observed trends if similar systems are being compared. This issue will be discussed later on.

Table 4. Excess enthalpies of the acetophenone + {N-methylformamide or N,N-dimethylformamide, or N,N-dimethylacetamide, or N-methyl-2-pyrrolidone} mixtures

| $x_1$ | $H^E/$ J·mol$^{-1}$ | $x_1$ | $H^E/$ J·mol$^{-1}$ | $x_1$ | $H^E/$ J·mol$^{-1}$ |
|---|---|---|---|---|---|
| Acetophenone (1) + N-methylformamide (2), $T$ = 293.15 K | | | | | |
| 0.0300 | 28.1 | 0.3473 | 366.3 | 0.6707 | 474.2 |
| 0.0600 | 67.3 | 0.3704 | 383.1 | 0.7007 | 466.7 |
| 0.0899 | 104.7 | 0.4004 | 403.0 | 0.7307 | 450.7 |
| 0.1199 | 140.6 | 0.4305 | 420.8 | 0.7606 | 422.4 |
| 0.1499 | 174.4 | 0.4605 | 436.5 | 0.7907 | 381.1 |
| 0.1799 | 206.5 | 0.4906 | 449.7 | 0.8206 | 350.5 |
| 0.2099 | 236.8 | 0.5206 | 460.6 | 0.8506 | 311.3 |
| 0.2399 | 265.4 | 0.5507 | 469.3 | 0.8805 | 283.9 |
| 0.2698 | 292.6 | 0.5807 | 475.0 | 0.9104 | 215.2 |
| 0.2998 | 317.8 | 0.6107 | 471.7 | 0.9402 | 160.8 |
| 0.3298 | 344.1 | 0.6407 | 475.7 | 0.9700 | 88.0 |
| Acetophenone (1) + N-methylformamide (2), $T$ = 308.15 K | | | | | |
| 0.0300 | 45.5 | 0.3704 | 405.6 | 0.7008 | 457.4 |
| 0.0600 | 88.9 | 0.4004 | 423.3 | 0.7308 | 444.2 |
| 0.0899 | 130.0 | 0.4304 | 438.7 | 0.7607 | 425.1 |
| 0.1199 | 169.3 | 0.4605 | 452.2 | 0.7907 | 401.5 |
| 0.1498 | 206.6 | 0.4905 | 463.2 | 0.8206 | 372.2 |
| 0.1799 | 242.0 | 0.5206 | 471.4 | 0.8506 | 329.6 |
| 0.2098 | 275.4 | 0.5507 | 477.1 | 0.8805 | 295.3 |
| 0.2398 | 306.8 | 0.5807 | 479.8 | 0.9105 | 233.6 |
| 0.2698 | 336.2 | 0.6107 | 479.5 | 0.9402 | 179.2 |
| 0.2998 | 363.6 | 0.6407 | 475.8 | 0.9701 | 102.3 |
| 0.3403 | 385.5 | 0.6707 | 468.9 | | |

Acetophenone (1) + N,N-dimethylformamide (2), *T* = 293.15 K

| | | | | | |
|---|---|---|---|---|---|
| 0.0300 | -14.5 | 0.3598 | -125.6 | 0.6989 | -115.7 |
| 0.0600 | -27.7 | 0.3898 | -130.0 | 0.7288 | -109.3 |
| 0.0899 | -41.0 | 0.4297 | -134.3 | 0.7604 | -101.5 |
| 0.1199 | -55.1 | 0.4597 | -135.7 | 0.7904 | -92.6 |
| 0.1499 | -68.3 | 0.4896 | -136.4 | 0.8204 | -82.9 |
| 0.1799 | -80.0 | 0.5195 | -136.2 | 0.8502 | -72.4 |
| 0.2098 | -89.9 | 0.5494 | -135.0 | 0.8802 | -59.5 |
| 0.2398 | -98.8 | 0.5793 | -133.2 | 0.9102 | -46.2 |
| 0.2698 | -107.1 | 0.6092 | -130.2 | 0.9401 | -32.2 |
| 0.2998 | -114.2 | 0.6392 | -126.2 | 0.9701 | -17.0 |
| 0.3298 | -120.4 | 0.6691 | -121.6 | | |

Acetophenone (1) + N,N-dimethylformamide (2), *T* = 308.15 K

| | | | | | |
|---|---|---|---|---|---|
| 0.0299 | -15.9 | 0.3598 | -119.5 | 0.7005 | -108.6 |
| 0.0599 | -30.6 | 0.3898 | -122.8 | 0.7304 | -102.4 |
| 0.0899 | -44.2 | 0.4303 | -125.7 | 0.7604 | -95.4 |
| 0.1199 | -56.9 | 0.4604 | -127.2 | 0.7904 | -87.4 |
| 0.1499 | -68.4 | 0.4904 | -127.9 | 0.8204 | -78.4 |
| 0.1799 | -78.8 | 0.5204 | -127.7 | 0.8503 | -68.0 |
| 0.2098 | -88.1 | 0.5504 | -126.6 | 0.8802 | -56.6 |
| 0.2398 | -96.4 | 0.5805 | -124.7 | 0.9102 | -44.3 |
| 0.2698 | -103.7 | 0.6104 | -122.0 | 0.9401 | -31.0 |
| 0.2998 | -109.9 | 0.6405 | -118.3 | 0.9700 | -16.2 |
| 0.3298 | -115.2 | 0.6705 | -113.9 | | |

Acetophenone (1) + N,N-dimethylacetamide (2), *T* = 293.15 K

| | | | | | |
|---|---|---|---|---|---|
| 0.0300 | -50.4 | 0.3597 | -359.5 | 0.6673 | -355.3 |
| 0.0599 | -111.8 | 0.3897 | -369.9 | 0.6975 | -339.1 |
| 0.0898 | -149.2 | 0.4197 | -377.7 | 0.7277 | -320.1 |
| 0.1198 | -183.0 | 0.4469 | -386.2 | 0.7578 | -298.1 |
| 0.1498 | -214.2 | 0.4569 | -388.1 | 0.7881 | -273.0 |
| 0.1798 | -242.8 | 0.4869 | -391.8 | 0.8183 | -244.7 |
| 0.2097 | -268.7 | 0.5169 | -392.7 | 0.8485 | -213.3 |
| 0.2397 | -292.1 | 0.5470 | -390.9 | 0.8788 | -178.4 |
| 0.2697 | -312.9 | 0.5770 | -386.3 | 0.9091 | -139.9 |
| 0.2997 | -331.0 | 0.6071 | -378.8 | 0.9394 | -97.4 |
| 0.3297 | -346.5 | 0.6372 | -368.5 | 0.9697 | -49.9 |

Acetophenone (1) + N,N-dimethylacetamide (2), *T* = 308.15 K

| | | | | | |
|---|---|---|---|---|---|
| 0.0302 | -42.1 | 0.3919 | -353.0 | 0.6705 | -337.1 |
| 0.0604 | -81.8 | 0.4219 | -363.3 | 0.7005 | -321.2 |
| 0.0906 | -118.8 | 0.4339 | -366.5 | 0.7306 | -302.5 |
| 0.1207 | -152.9 | 0.4484 | -370.7 | 0.7605 | -281.1 |
| 0.1509 | -184.8 | 0.4604 | -372.7 | 0.7906 | -256.8 |
| 0.1811 | -214.2 | 0.4905 | -375.6 | 0.8206 | -229.7 |
| 0.2113 | -240.9 | 0.5205 | -375.9 | 0.8505 | -199.7 |
| 0.2414 | -264.9 | 0.5504 | -373.6 | 0.8803 | -166.9 |
| 0.2715 | -286.2 | 0.5805 | -368.5 | 0.9103 | -130.5 |
| 0.3016 | -304.9 | 0.6105 | -360.8 | 0.9402 | -91.2 |
| 0.3317 | -323.8 | 0.6405 | -350.3 | 0.9701 | -47.9 |
| 0.3618 | -339.9 | | | | |

Acetophenone (1) + N-methyl-2-pyrrolidone (2), $T$ = 308.15 K

| | | | | | |
|---|---|---|---|---|---|
| 0.0299 | -88.9 | 0.3597 | -683.7 | 0.7003 | -616.0 |
| 0.0599 | -171.7 | 0.3897 | -703.7 | 0.7303 | -579.9 |
| 0.0898 | -248.9 | 0.4602 | -720.5 | 0.7604 | -537.1 |
| 0.1198 | -320.2 | 0.4902 | -722.0 | 0.7903 | -487.7 |
| 0.1498 | -385.5 | 0.5202 | -719.0 | 0.8203 | -435.2 |
| 0.1797 | -444.9 | 0.5502 | -713.4 | 0.8503 | -375.8 |
| 0.2097 | -499.1 | 0.5803 | -704.0 | 0.8803 | -310.7 |
| 0.2397 | -547.4 | 0.6103 | -689.8 | 0.9102 | -241.8 |
| 0.2697 | -590.0 | 0.6403 | -670.4 | 0.9401 | -162.5 |
| 0.2997 | -627.0 | 0.6703 | -645.9 | 0.9701 | -83.7 |
| 0.3297 | -658.3 | | | | |

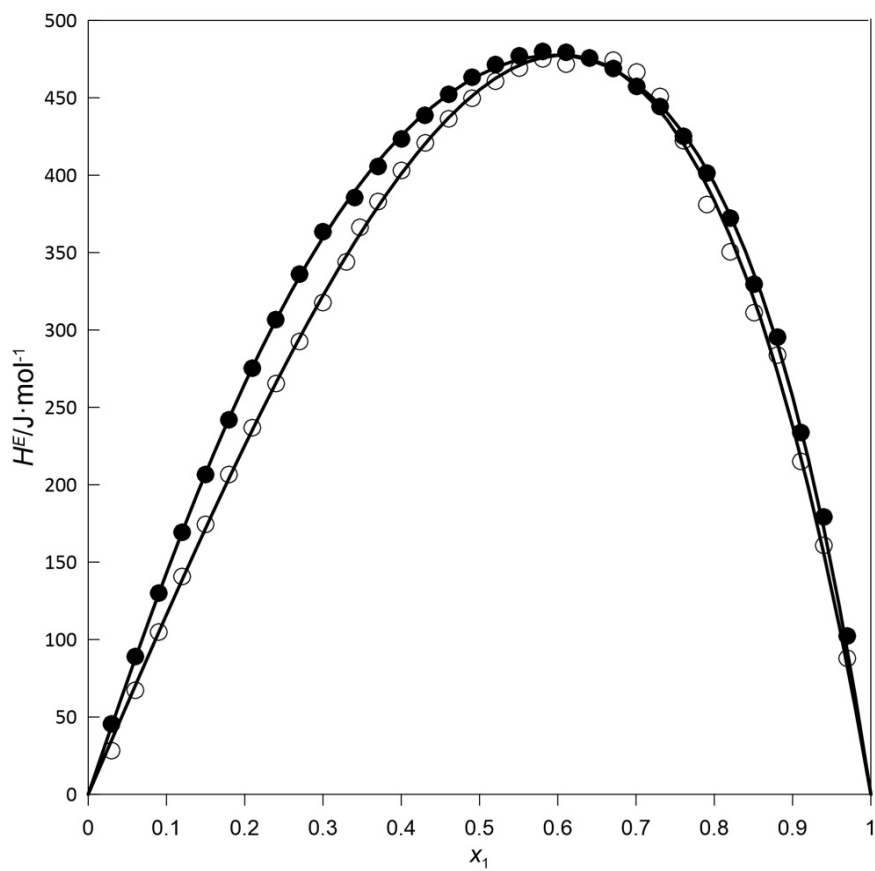

Fig. 4

Excess enthalpies for the acetophenone + N-methylformamide system. Experimental: (○) 293.15 K, (●) 308.15 K, solid lines are calculated by eqn. (6) with the parameters given in Table 6.

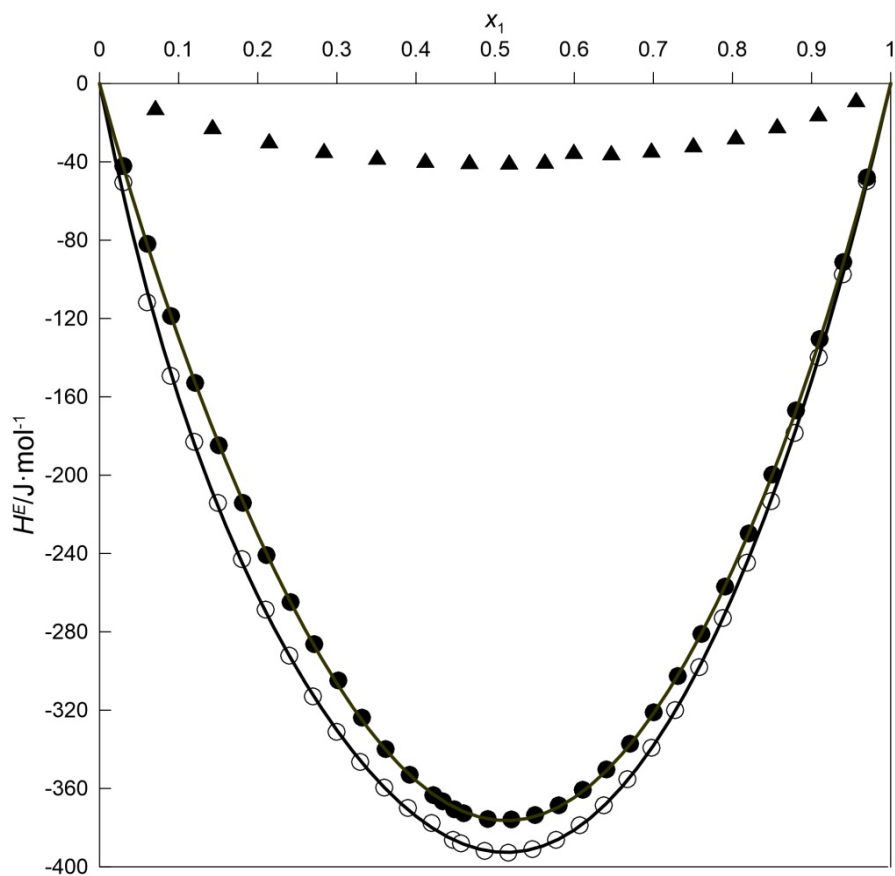

Fig. 5

Excess enthalpies for the acetophenone + N,N-dimethylacetamide system. Experimental: (○) 293.15 K, this work, (●) 308.15 K, this work , (▲) 298.15 K, Sekhar et al. [5], solid lines are calculated by eqn. (6) with the parameters given in Table 6.

## 4. Thermodynamic description

For a binary system with a pure solid (1) being in equilibrium with liquid mixture and if (i) no transition points in the solid phase appear, and (ii) the enthalpy of melting is independent of temperature, the solubility is represented by the following equation

$$R \ln[x_1 \gamma_1(x_1, T)] = -\Delta H_{m1} \left( \frac{1}{T} - \frac{1}{T_{m1}} \right) \tag{1}$$

where $x_1, \gamma_1$, $\Delta H_{m1}, T_{m1}$ are mole fraction, activity coefficient, enthalpy and temperature of melting pertaining to the solute, respectively. In this equation, the solute activity coefficient as a function of concentration and temperature is of crucial significance as it in fact determines the above equation. It results directly from the differences between molecular interactions and entropies occurring in a solution. The solute activity coefficients can be treated as experimental if calculated using the above equation, provided that other parameters are measured or known. From the practical point of view, their dependence can have a fully correlational character and does not have to be related to the excess

Gibbs energy if adjusted to the experimental solubility data only. However, the latter approach is frequently used as it leads to the solvent activity coefficient thus giving a chance to obtain a full thermodynamic description of the system. Unfortunately, this way the excess Gibbs energy cannot be determined unambiguously as activity coefficients for non-isothermal data depend also on the excess enthalpy according to the Gibbs-Duhem equation

$$x_1 d \ln \gamma_1 + x_2 d \ln \gamma_2 = -\frac{H^E}{RT^2} dT \qquad (2)$$

It means, that if a well-defined model for the $G^E(T,x_1)$ is adjusted to the experimental SLE data, both the solvent activity coefficient and excess enthalpy are determined being mutually interrelated. And on the other hand, for the same set of the experimentally derived solute activity coefficients and using a perfectly adjusted $G^E$ model, one can obtain various values of the second activity coefficient depending on the excess enthalpies assumed or expected. Two limiting cases can be considered: (i) if $G^E/T \neq f(T, x_1 = \text{const})$ then $G^E = -TS^E$, $\gamma_1 \neq f(T, x_1 = \text{const})$ and $H^E = 0$; (ii) $G^E \neq f(T, x_1 = \text{const})$ then $G^E = H^E$, $c_p{}^E = 0$ and $RT\ln\gamma_1 \neq f(T, x_1 = \text{const})$. The practical problem in correlation results from the fact that the adjusted activity coefficient function is directly dependent on solubility which is an implicit function of temperature. Both dependencies on temperature and composition cannot be separated without additional information, i.e. a particular form of the excess Gibbs energy function or experimental excess enthalpies.

For the general (but isobaric) case, the interrelations between activity coefficients and excess Gibbs energy has the following form

$$\ln \gamma_1 = Q + x_2 \left(\frac{dQ}{dx_1}\right) + \frac{x_2 H^E}{RT^2} \left(\frac{dT}{dx_1}\right) \qquad (3a)$$

$$\ln \gamma_2 = Q - x_1 \left(\frac{dQ}{dx_1}\right) - \frac{x_1 H^E}{RT^2} \left(\frac{dT}{dx_1}\right) \qquad (3b)$$

where $Q = G^E/RT$ and all differentiations should be performed along the solubility curve. Using this representation, eqn. (1) can be re-written as below

$$R \ln x_1 + R \left[Q + x_2 \left(\frac{dQ}{dx_1}\right) + \frac{x_2 H^E}{RT^2} \left(\frac{dT}{dx_1}\right)\right] = -\Delta H_{m1} \left(\frac{1}{T} - \frac{1}{T_{m1}}\right) \qquad (4)$$

with the analogous form for the component 2 forming the solid phase.

To accomplish a determination of the solute and solvent activity coefficients based on the above equation, three conditions should be fulfilled: (i) the $G^E/RT$ as an explicit function of solubility only must be assumed; (ii) excess enthalpy at the optional temperature and concentration must be known; (iii) the derivative $\left(\frac{dT}{dx_1}\right)$ along the solubility curve should be determined.

## 5. Calculations

For the $Q$-function the simplest power expansion in the form of the Redlich-Kister equation was used, i.e.

$$G^E/RT = x_1 x_2 \sum_{i=0}^{n} a_i (x_1 - x_2)^i \qquad (5)$$

with the $a_i$ parameters being independent both of $x_1$ and $T$. Similarly, the Redlich-Kister equation was used to correlate experimental excess enthalpies. To account for the impact of temperature, a linear dependence of their parameters was assumed

$$H^E = x_1 x_2 \sum_{i=0}^{n}(b_{i0} + b_{i1}T/T_0)(x_1 - x_2)^i \tag{6}$$

with the reference temperature $T_0$ equal to 298.15 K.

The power expansion has been chosen for both representations to minimize a possible influence of the assumed model used to correlate the data. The number of the expansion terms ($n$+1) was kept at the lowest possible level, at the same time preserving consistency with the experimental uncertainty. Only two parameters were assumed for the temperature dependence in eqn. (6) as the excess enthalpy data measured at two temperatures do not justify a higher number of parameters.

If the excess enthalpies for the optional temperature and concentration are known, the only unknown is the $Q$-function which can be adjusted to the experimental solubility data and then used to calculate activity coefficients using eqns. (3a) and (3b). Such an approach can be difficult to perform practically, as the final solubility equation is a nested implicit equation because of the derivative $\left(\frac{dT}{dx_1}\right)$ which can be calculated provided that eqn. (4) is solved with respect to the temperature. Formally, an iterative procedure with a numerical differentiation could be used. Alternatively we propose to calculate this derivative separately, using a semi-empirical equation smoothing the experimental solubility data by a sufficiently flexible equation. For this purpose the equation

$$\ln x_i = \frac{A}{T} + B \ln T + C \tag{7}$$

was used. It is sometimes called as the Apelblat equation[20] although this form was applied earlier as it clearly results from the integration of the equilibrium condition expressed in the exact differential form. The values of the adjusted parameters with the standard deviations of temperature are given in Table 5. It should be noted that no physical significance can be assigned to its parameters. Their values are strongly correlated and they should be treated as the regression parameters only. We used this equation only to estimate the $\left(\frac{dT}{dx_1}\right)$ derivative.

Table 5. Correlation of the solubility data by means of eqn. (7)

| System | $10^{-3} \cdot A$ /K | $B$ | $C$ | $\sigma^a$ /K |
|---|---|---|---|---|
| Acetophenone + | 10.139 | 52.410 | -332.20 | 0.36 |
| N-methylformamide | -39.810 | -143.20 | 949.13 | 0.18 |
| Acetophenone + | 1.2049 | 11.726 | -70.665 | 0.80 |
| N,N-dimethylformamide | | | | |
| Acetophenone + | 2.0370 | 13.980 | -86.384 | 0.42 |
| N,N-dimethylacetamide | 9.4752 | 42.919 | -274.93 | 0.55 |
| Acetophenone + | 4.3966 | 21.844 | -139.07 | 0.32 |
| N-methyl-2-pyrrolidone | 30.636 | 131.02 | -845.92 | 0.56 |

$^a$ Standard deviation $\sigma = \left[\frac{1}{m}\sum_{i=1}^{m}\left(T_i^{exp} - T^{calc}\left(x_{1i}^{exp}\right)\right)^2\right]^{1/2}$

Finally, values of the parameters of eqn. (5) were adjusted to the experimental solubility data through the minimization of the following objective function

$$F(a_0, a_1, \dots, a_n) = \sum_{i=1}^{m} \left[ T_i^{exp} - T^{calc}\left(x_{1i}^{exp}; a_0, a_1, \dots, a_n\right) \right]^2$$

where superscripts *exp* and *calc* denote experimental and calculated value of a parameter corresponding to the *ith* experimental point. The $T^{calc}$ value was calculated by a solution of the non-linear equation with respect to $T$ with the $H^E$ and $\left(\frac{dT}{dx_1}\right)$ expressed as a function of temperature and composition derived from eqn. (6) and (7).

Table 6 shows results for the excess enthalpy and the SLE correlation performed by the manner described above together with the calculated solvent activity coefficients and values of the adjusted parameters of eqn. (5) and (6). The solid lines in Figures 2-5 represent dependencies calculated using the adjusted models.

Table 6. Correlation of the excess enthalpy and solid-liquid equilibrium data by the Redlich-Kister equation (5, 6)

| System Acetophenone + | $i$ | $H^E$ | | | SLE | |
| --- | --- | --- | --- | --- | --- | --- |
| | | $b_{i0}$ /J·mol$^{-1}$ | $b_{i1}$ /J·mol$^{-1}$ | $\sigma^a$ /J·mol$^{-1}$ | $a_i$ | $\sigma^b$ /K |
| N-methylformamide | 0 | 1076.4 | 757.08 | 4.3 | 1.3364 | 0.39 |
| | 1 | 6043.8 | -5339.8 | | 0.19356 | |
| | 2 | -6391.3 | 6732.5 | | 0.24819 | |
| | 3 | -6397.0 | 6592.1 | | 0.17192 | |
| N,N-dimethylformamide | 0 | -1232.1 | 696.59 | 0.40 | -0.2309 | 0.47 |
| | 1 | -116.36 | 121.66 | | 0.32348 | |
| | 2 | 935.54 | -951.98 | | -0.23234 | |
| | 3 | -859.33 | 816.64 | | | |
| N,N-dimethylacetamide | 0 | -2841.8 | 1294.0 | 1.3 | -0.29744 | 0.51 |
| | 1 | 67.017 | -151.88 | | 0.13598 | |
| | 2 | -3592.1 | 3555.6 | | | |
| | 3 | 4781.0 | -4635.7 | | | |
| | 4 | -1936.3 | 1711.4 | | | |
| N-methyl-2-pyrrolidone | 0 | -2897.6 | 0 | 2.2 | -0.91458 | 0.53 |
| | 1 | 80.303 | 0 | | 0.60981 | |
| | 2 | -425.12 | 0 | | -0.11981 | |
| | 3 | -43.110 | 0 | | | |
| | 4 | 446.66 | 0 | | | |

[a,b] Standard deviations: (a) $\sigma = \left[ \frac{1}{m} \sum_{i=1}^{m} \left( H_i^{E\,exp} - H^{E\,calc}(T_i, x_{1i}^{exp}) \right)^2 \right]^{1/2}$; (b) $\sigma = \left[ \frac{1}{m} \sum_{i=1}^{m} \left( T_i^{exp} - T^{calc}\left(x_{1i}^{exp}\right) \right)^2 \right]^{1/2}$

Prediction of the measured data can be performed by means of any group contribution model provided that necessary group parameters are known. By no means it is the UNIFAC approach which, involving some modifications, is mainly applied to predict thermodynamic properties under low and moderate pressures. Its main practical advantages lies in a comprehensive and relatively easily accessible matrix of the group parameters. There appear, however, a problem with a proper group definition. If the idea of the UNIFAC-group assignment is to be followed, the carbonyl group adjacent to the aromatic ring should be recognized as a separate group although such a group was not defined yet or its parameters still are undetermined. It should be noted that NMF and DMF in the UNIFAC approach are treated as distinct groups and each one is equivalent to a whole molecule. It seems that division into smaller groups would be ineffective, but such procedure, however, disagrees with the very idea of the solution-of-groups approach.

To make possible the UNIFAC prediction we took the parameters of the CH3CO (aliphatic) group as a possible estimate for the carbonyl group adjacent to the aromatic ring – see Table 7. The modified UNIFAC model [21] was used with the group parameters developed by the NIST group [22]. Such approach  certainly is a crude simplification, nevertheless we believe that it can credibly estimate at least differences between properties of various systems with the same one component (i.e. acetophenone).

Table 7. The NIST-Modified UNIFAC group assignment[22]

| UNIFAC group | Group number [22] | Number of groups in a molecule of | | | | |
|---|---|---|---|---|---|---|
| | | Acetophenone | NMF | DMF | DMA | NMP |
| CH3 | 1 | | | | 1 | |
| ACH | 9 | 5 | | | | |
| AC | 10 | 1 | | | | |
| CH3CO | 18 | 1 | | | | |
| DMF | 72 | | | 1 | | |
| c-CH2 | 78 | | | | | 3 |
| c-CON-CH3 | 86 | | | | | 1 |
| HCONHCH3 | 93 | | 1 | | | |
| CON(CH3)2 | 101 | | | | 1 | |

## 6. Results and discussion

All the solid-liquid equilibria form simple eutectics systems (Table 3, Fig. 2 and 3) with the eutectic temperature varying between 207.2 K (for acetophenone + DMF) and 261.6 K (for acetophenone + NMF). The values of the Redlich-Kister parameters adjusted to the data together with the standard deviations are given in Table 6. The model with two, three or four parameters is able to describe solubility curve with the standard deviations in the range of 0.4-0.5 K what agrees well with the experimental uncertainty. For the systems with NMF, solubilities are lower than ideal ones (positive deviations from ideality) while for remaining solutions they are higher (negative deviations from ideality). The isobaric vapor pressures measured by Vittal Prasad et al. [4] for the acetophenone +
DMF system agrees only qualitatively with our solid-liquid equilibria although a direct comparison is
not possible because of temperature differences of about 200 K between both data sets. Absolute
values of our excess Gibbs energies are about four times lower than calculated from the vapor-liquid
equilibrium measurements.

Excess enthalpies were measured at two temperatures – 293.15 K and 303.15 K except for the
acetophenone + NMP system, for which the measurements were performed at 303.15 K (Table 4, Fig.
4-6). The observed heat effects are exothermic for the systems with DMF, DMA and NMP and
endothermic for the acetophenone + NMF. The latter excess enthalpy function is unsymmetrical with
the maximum shifted to higher concentrations of acetophenone while for remaining mixtures it is
almost symmetrical. The Redlich-Kister equation with the parameters being linearly dependent on
temperature is able to describe experimental data with the standard deviations close to the
estimated uncertainty. The number of adjustable parameters per each system was found to be eight
or ten for over sixty experimental points (Table 6).

The temperature dependence of the excess enthalpy enables to estimate excess heat capacity by
differentiation with respect to temperature. It must be recalled, however, that an error of such
procedure is high and it can be regarded as a crude estimation only. The calculated excess heat
capacities are drawn in Fig. 6. Their irregular shape can merely result from the bias introduced by the
correlation equation applied, although the W-shape of the excess heat capacity for some polar
systems was observed and has been extensively studied [23].

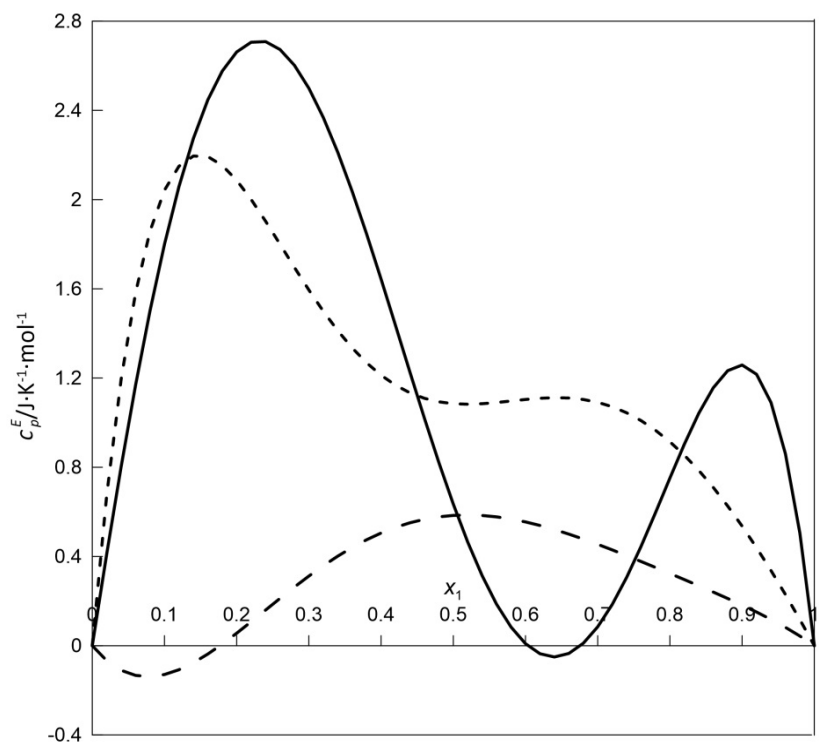

Fig. 6

Estimated excess heat capacities for the following systems of acetophene + : (solid) N-
methylformamide, (small dash) N,N-dimethylacetamide, (long dash) N,N-dimethylformamide.

Among the systems studied, that of acetophenone + N-methylformamide turned out to be outstanding as it is the only mixture exhibiting positive deviations from ideality. Such a behavior can be explained by possible self-association of N-methylformamide molecules which is suppressed if a solvent is added. In the NMF multimers, which are being created, the hydrogen atom in the amine group acts as a donor and joins oxygen atom from the carbonyl group. It is supposed that this self-association is mainly responsible for the extremely high dielectric constant of N-methylformamide[24]. An analogous association is not possible between N,N-dimethylamides and N-methylpyrrolidine molecules. One could expect another mechanism of self-association in both formamides, i.e. NMF and DMF, in which hydrogen adjacent to the carbon of carbonyl group is a donor while amine electron pair serves as a acceptor. However, it does not look probable as the system of acetophenone with DMF exhibits significant negative deviations from ideality.

There are additional thermodynamic arguments supporting suggestion that in the system acetophenone + N-methylformamide, the self-asscociation of the latter component predominates. That is positive and unsymmetrical excess enthalpy, and positive and relatively high excess heat capacity. They are similar to the same properties of an alkanol + *n*-alkane mixtures which represent model systems with one strong self-associated component – see for example[25]·[26].

Almost for all the systems entropic contribution to excess Gibbs energy is significant although (the exception is the system with DMF for which it seems to be meaningless) but it is the enthalpic contribution which determines the character (i.e. positive or negative) of deviations from ideality. It seems that an approximate explanation of observed properties can be restricted to mutual interactions only.

The mixtures of acetophenone with {DMF or DMA , or NMP} show negative deviations from ideality. Such systems are relatively rare and their properties usually are explained by dominant attractive interactions between unlike molecules which are stronger than interactions between like molecules. What kind of specific interactions between acetophenone and an amide can invoke negative deviations from ideality? The answer is not easy. As the system of DMF with acetone exhibits positive and relatively low excess enthalpies [27], it seems that the aromatic ring has a greater impact on this property. Indeed, excess enthalpies for a series of systems: benzene + {DMF, +DMA, + NMP) decrease from low and positive [28], through negative [28], down to minimum value of about -500 J·mol$^{-1}$ [29]. However, it is not clear what particular strong specific interactions can occur.

Interesting observations can be drawn from the application of the modified UNIFAC model [21]. As it was noted in the previous section, the parameters of the aliphatic CH3CO group were used to account for the carbonyl group in the acetophenone molecule and only qualitative prediction is expected. If only character of deviations from ideality is concerned, properties of majority of systems are correctly predicted. The only exception is solid-liquid equilibria for the acetophenone + DMA system for which the UNIFAC model erroneously predicts solubility lower than ideal (Fig. 3). On the other hand, the excess enthalpy for the same system is predicted almost perfectly. For the system with DMF, if we neglect the predicted S-shaped dependence, the calculated excess enthalpies are mainly negative and with relatively low absolute values, what generally agrees with reality. A strong disagreement can be observed for the acetophenone + NMP mixture for which again S-shaped dependence of almost meaningless excess enthalpies is predicted against very low and negative experimental values (Fig. 7).

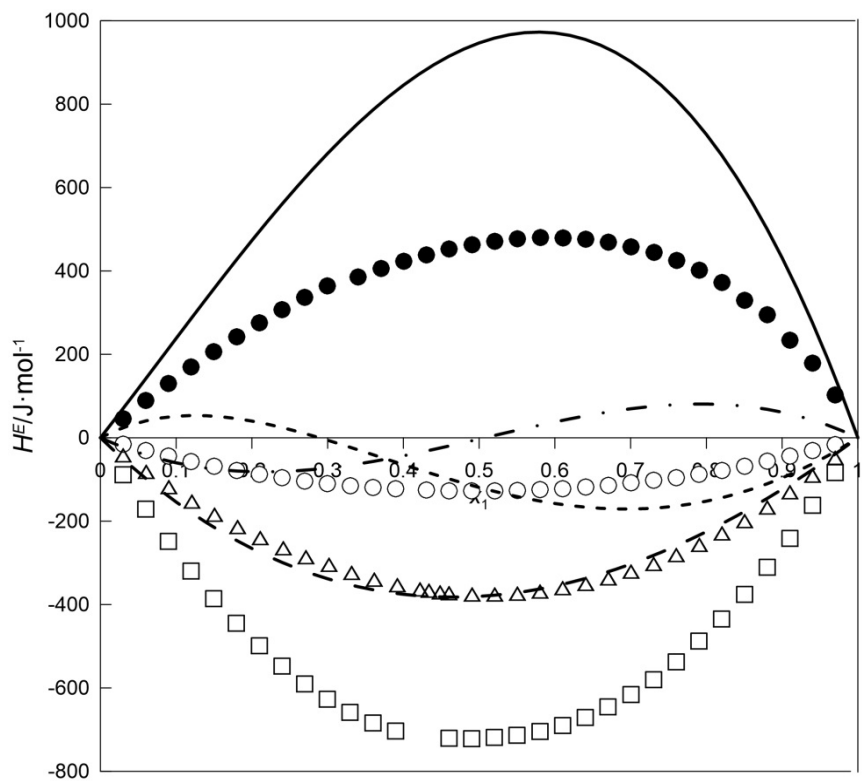

Fig. 7

Experimental excess enthalpies (symbols) and predicted by the modified UNIFAC model (lines) at 308.15 K for the system acetophenone + , (●, solid) N-methylformamide, (○, small dash) N,N-dimethylformamide, (Δ, long dash) N,N-dimethylacetamide, (□, dash + dot) N-methyl-2-pyrrolidone.

The predicted data suggest that excess enthalpies for the acetophenone + DMA mixtures are lower (i.e. more negative) than for the system with DMF. In spite of obvious weakness of such reasoning we recognize this observation as a partial support of our opinion that our data for the former system are more credible than those of Sekhar et al.[5] Similar relation is observed for experimental excess enthalpies for analogous systems in which acetophenone is replaced by propiophenone [30]. The same character of residuals may be noted if the excess enthalpies of DMF + benzene (positive and close to zero) are compared with those of DMA + benzene (negative)[28].

## 8. Conclusions

The solid-liquid equilibrium and excess enthalpy data for solutions of acetophenone in {N-methylformamide,  or N,N-dimethylformamide, or N,N-dimethylacetamide, or N-methyl-2-pyrrolidone – NMP were measured and successfully correlated by means of the Redlich-Kister equation with standard deviations close to experimental uncertainty. The acetophenone + NMF system exhibits strong positive deviations from ideality what manifests itself in positive excess enthalpies and decreased solubilites of acetophenone in comparison with the ideal solubility. The remaining systems reveal opposite properties as both negative deviations from ideality and negative excess enthalpies are observed. A larger extent of negative deviations from ideality was observed in the acetophenone + DMA system than in the acetophenone + DMF system. This phenomenon is not understood and requires further examinations. It is not unique, however, as a similar property can be found in different systems containing DMA and DMF. An application of the modified UNIFAC model show disadvantages of the method if the prediction even slightly goes beyond the space of defined group parameters. For majority of systems the prediction is qualitative or semi-qualitative but for single systems it may be unacceptable and hardly can be found a clear hint what accuracy can be expected for a particular mixture.

## 9. Acknowledgements

This work was financially supported by Warsaw University of Technology. One of us (A. Cobos) acknowledges support of Spanish government in the framework of xxxxx.